\documentclass[format=acmsmall]{acmart}
\usepackage[utf8]{inputenc}
\usepackage{multirow}
\usepackage{quoting}
\usepackage{makecell}

\acmJournal{CSUR}

\title{Blockchains and Self-Sovereign Identities Applied to Healthcare Solutions: A Systematic Review}

\author{Alexandre Siqueira}
\email{alexandre.siqueira@unifesp.br}
\author{Arlindo F. da Conceição}
\email{arlindo.conceicao@unifesp.br}
\affiliation{%
  \institution{Federal University of São Paulo}
  \department{Instituto de Ciência e Tecnologia}
  \city{São José dos Campos}
  \state{SP}
  \country{Brazil}
}
\author{Vladimir Rocha}
\email{vladimir.rocha@ufabc.edu.br}
\affiliation{%
  \institution{Federal University of ABC}
  \department{Centro de Matemática, Computação e Cognição}
  \city{Santo André}
  \state{SP}
  \country{Brazil}
}

\acmVolume{1}
\acmNumber{1}
\acmArticle{1}
\acmYear{2021}
\acmMonth{4}

\acmSubmissionID{}
\received{April 2021}
\acmDOI{}
\acmPrice{}

\ccsdesc[500]{Applied computing~Health care information systems}
\ccsdesc[500]{Security and privacy~Privacy protections}
\ccsdesc[300]{Computer systems organization~Peer-to-peer architectures}
\ccsdesc[300]{General and reference~Surveys and overviews}

\keywords{privacy, electronic health records, self-sovereign identity, SSI, blockchain, interoperability}

\begin{abstract}
Self-Sovereign Identity (SSI), a Blockchain-based technology for digital identity management, is a promising concept for handling health data. It could represent a step forward in empowering users, granting them control over their data. This work conducts a systematic literature review to investigate state-of-the-art measures based on SSI and Blockchain technologies for dealing with electronic health records (EHRs), identifying gaps, and determining the key questions for future research. As a result, this review shows a growing interest in Blockchain methods to handle EHRs, but few works consider using the self-sovereign identity approaches. The results obtained in this work also suggest that: Blockchain technologies provide a viable alternative to deliver EHR solutions such as patient monitoring, healthcare data trading, and prescription control; consolidated Blockchain technologies are the preferred core components of most effective strategies; keeping raw health data off-chain helps to create scalable solutions; health data standards make searching medical records in Blockchain structures feasible; Smart Contracts are essential components of Blockchain-based EHR solutions; the concepts of data ownership and Self-Sovereign Identity have been neither adequately defined nor employed in the health context.
\end{abstract}

\begin{document}
\maketitle

\section{Introduction}

In the digital era, our data is currency. Technology companies like Google and Facebook are among the most valued companies in the world, and make much of their money by selling to other companies targeted advertisement opportunities based on the data and behavior of people that use their platforms and apps~\cite{eMarketer_2019}. In exchange, the users have free services such as e-mail accounts, digital storage for their photos, and the opportunity to interact with other people through social networks. All kinds of companies follow this trend: online retailers like Amazon and Alibaba are profiling and scoring their customers on the basis of their browsing and purchasing history; ride-sharing companies like Uber and Lyft collect user's geolocation data about their users to optimize their fleet distribution and stimulate their partner drivers with more attractive fares. Harcourt summarizes this situation by stating that~\cite{Harcourt_2015}: ``today every single digital trace can be identified, stored, and aggregated to constitute a composite sketch of what we like, whom we love, what we read, how we vote, and where we protest.''

This ``currency exchange'' can be viewed as legitimate, for as long as there is consent. Laws to protect personal data such as GDPR~\cite{gdpr_2016} and its local copycats have emerged as a way of regulating this exchange and making it fairer, establishing principles of use (purpose, need, free access, right to be forgotten, transparency among others) and imposing penalties for abusive conduct.

In light of this, privacy protection laws are essential instruments for control and regulation but insufficient for establishing the ownership of data as a personal right. This limitation is observed in the following clauses:

\begin{itemize}
    \item \textbf{Willingness to disclose personal data:} some people feel comfortable sharing their data in exchange for access to content, discounts, rewards, and other benefits~\cite{mazurek_2019}. In this regard, privacy protection laws impose a barrier for companies that want to provide services to individuals that regard data as a personal commodity and are willing to make a trade-off.
    \item \textbf{Digital surveillance:} The Cambridge Analytica scandal~\cite{nytimes_2018} revealed that companies could neglect their responsibilities as custodians of customers' private data and thus enable third-parties to take advantage of their digital tracking mechanisms.
\end{itemize}

Personal health records reveal a conflict between privacy protection laws and the ability to share personal information. Throughout life, people interact with doctors and healthcare service providers countless times, either for routine appointments or medical treatment, clinical information is obtained that could constitute a significant batch of medical records for future use. However, a single patient might have his/her medical data spread across several healthcare service providers, and thus create siloed databases that, in the current state, are useless for clinical application outside those silos. Added to this scenario, the increasing adoption of wearable devices, for collecting information about health and lifestyle, is creating a new patient-generated health data silo.

The fragmented nature of personal medical records prevents someone from having a unique and usable person's health track record that could help physicians make an accurate diagnosis. New methods are needed for creating a decentralized data structure, that is controlled by the users and allows them to maintain their medical records. This appears to be the pathway to enabling health providers to store medical information while granting data ownership to the rightful owners - the patients.

The dilemma of data ownership can be overcome by the use of Digital Identities, particularly the Self-Sovereign Identities (SSI) concept: this approach allows the user to take control of his/her data, granting permissions to which data could be accessed and by whom, and revoking this access at any time~\cite{Allen_2016}. Kassab \textit{et al.}~\cite{kassab_2019} and Houtan \textit{et al.}~\cite{Houtan_2020} have investigated the use and benefits of Decentralized Ledger Technologies (DLT), such as Blockchain. This kind of technology could be a means of confronting the challenge of sharing and securing sensitive medical information among healthcare parties, as well as ensuring patients maintain sovereignty over their data.

This study investigates the state-of-the-art approaches for dealing with electronic health records by conducting a comprehensive Systematic Literature Review (SLR) of the adopted Blockchain technologies, privacy concerns, and interested parties. 

The study is structured as follows: Section~\ref{sec:background} provides an outline of the historical background to electronic health records and related work. Section~\ref{sec:method} defines the research questions and the methodology employed for carrying out the systematic literature review. Section~\ref{sec:results} describes the results of the review and conducts an analysis related to the research questions. Section~\ref{sec:challenges} mentions some of the challenges facing future research. Section~\ref{sec:conclusions} summarizes our findings and conclusions.

\section{Background and Related Work}
\label{sec:background}

This section explains the terms and concepts used in this review and also examines related work.

\subsection{Definitions}

The following definitions are designed to clarify main concepts and elucidate the objectives and findings of this study.

\hfill

\noindent \textbf{Electronic health records:} Hardin and Kotz~\cite{Hardin_2019} describe electronic health records (EHR) as:  ``real-time, patient-centered records that make information available instantly and securely to authorized users.'' It is the digital version of a patient's medical track record. 

\hfill

\noindent \textbf{Interoperability:} As defined by the IEEE Standard Computer Dictionary, interoperability is ``The ability of two or more systems or components to exchange information and to use the information that has been exchanged''~\cite{Ieee_1991}.

\hfill 

\noindent \textbf{Privacy:} It is stated in the Universal Declaration of Human Rights~\cite{Udhr_1948}: ``No one shall be subjected to arbitrary interference with his privacy, family, home or correspondence, nor to attacks upon his honour and reputation''. Our study employs the concepts of privacy as outlined by Solove~\cite{Solove_2008} and follows his approach by understanding privacy through its violations. This taxonomy groups activities that are harmful to privacy in the following categories:

\begin{itemize}
    \item \textbf{Information collection:} third-party entities collect information by means of surveillance (watching, listening to, or recording), and interrogation (questioning or probing);
    \item \textbf{Information processing:} the information is stored, kept, and used through aggregation (a combination of the data), identification (attributing the data to an individual), insecurity (lacking protection of the stored data), secondary use (using the data for a different purpose from what was originally intended), and exclusion (preventing people from knowing what data are stored about them);
    \item \textbf{Information dissemination:} the information is released or shared by the data holders in what involves some kind of a breach of confidentiality (breaking the agreement to keeping the information confidential), disclosure (revelation of a person's information), exposure (similar to disclosure but involving physical matters, such as nudity), increased accessibility (facilitating the exposure of information and accessibility), and blackmail (threatening to disclose the information);
    \item \textbf{Invasion:} third-party entities interfere with a person's private life using intrusion (with invasive acts) or decisional interference (involving intrusion by the government into people's private decisions).
\end{itemize}

\hfill

\noindent \textbf{Blockchain:} ``Blockchain is an ordered list of records linked together through a chain on blocks''~\cite{Wang_2019}. Following its advent in 2008, Nakamoto~\cite{Nakamoto_2008} defined Blockchain as the building block of a new payment system that would begin the cryptocurrency movement. Since then, the Blockchain concept has been adapted and applied to many other scenarios because of its unique features~\cite{Werbach_2018}:

\begin{itemize}
    \item Decentralized control (distributed)
    \item A shared view of truth (immutable)
    \item Collaboration across boundaries of organizations (transparent)
    \item The direct exchange of value through tokens (crypto-economic)
\end{itemize}

Decentralized control means that the information is shared among the participants.As a result, the information is available even when some of the participants are offline or leave the system. The concept of sharing means that all the participants have the same information, and that this information cannot be altered when added to the list. The collaboration means that, as the information is immutable, it can be audited without any fear that it will be altered. In addition, as the information is linked, it can be tracked. The direct exchange provides a reward to participants as an incentive for maintaining the system and keeping it running.

In the Blockchain, each block comprises a header and a list of information relative to the scenario which is applied. The header is responsible for linking the current block with a predecessor block in the chain. At the same time, the list stores the information \textit{per se}. The information, called ``transaction'', is generally a metadata of some key attributes which allow it to be identified. For example, in the financial sector, the metadata could be the accounts and the amount of money transferred between the clients. In the medical sector, the metadata could be the patient identifier with the doctors' prescription.

When a scenario-specific functionality is added (e.g., to check if the prescription has a valid date) the Blockchain allows a developer to implement and deploy it as a smart contract. A smart contract is a software that is executed by each Blockchain participant, to enforce business requirements of the information that is stored. It should be mentioned that, like the blocks in the chain, the smart contract is immutable and transparent. 

These features make Blockchains suitable for applications that depend on trust. For this reason, Blockchains are the technology that supports solutions ranging from cryptocurrencies to smart contracts, going through tracking systems for logistics as well as issuing digital identities.

\hfill

%\noindent \textbf{Self-sovereign identities:} {\color{blue}The self-sovereign identity (SSI) concept, defined by Allen~\cite{Allen_2016}, ensures that the users have full control of their digital identities, in which they can share them with consent, creating user autonomy. Also, a SSI must allow users to make digital claims, such as personal information, capabilities, opinions, among others.}

%In light of this, self-sovereign identities have evolved from previous ways of managing digital identities. {\color{blue}The levels set out by Allen~\cite{Allen_2016}, in increasing order of evolution, are: centralized, federated, user-centric, and self-sovereign identity.

%In the centralized, the identity is controlled by a single authority (e.g., login/password of a website). In the federated, the identity is controlled by multiple federated authorities (e.g., a social media account used in a shopping website). In the user-centric, the identity is controlled by an individual across multiple authorities without a federation (e.g., cryptocurrency wallets stored by the user). Finally, in the self-sovereign, the identity is controlled by an individual across any number of authorities (e.g., a digital wallet that stores several digital identities).
%}

\noindent \textbf{Self-sovereign identities:} From the discussions regarding digital identities and standards for verifiable credentials, Allen~\cite{Allen_2016} outlined, in 2016, the first draft of the concept he called "self-sovereign identities". Allen described an evolutionary timeline of digital identity management models, from the original centralized models, advancing through federated models, finally reaching his concept of self-sovereign identities:

\hfill

\begin{quoting}[rightmargin=0cm,leftmargin=4cm]
\footnotesize
\noindent
    \textit{``Self-sovereign identity is the next step beyond user-centric identity and that means it begins at the same place: the user must be central to the administration of identity. That requires not just the interoperability of a user’s identity across multiple locations, with the user’s consent, but also true user control of that digital identity, creating user autonomy.''} 
\end{quoting}

\hfill

To ensure the ownership of individuals over their identities, Allen describes ten principles self-sovereign identities must observe:

%{\color{blue}According to Allen, to achieve the last level (i.e., the SSI), it is necessary to follow ten principles, which ensure users will have total control over its identities and claims. The most related to this work are:

\begin{enumerate}
    \item \textbf{Existence:} Users must have an independent existence;
    \item \textbf{Control:} Users must control their identities;
    \item \textbf{Access:} Users must always have access to their identities and claims;
    \item \textbf{Transparency:} Systems and algorithms used to manage the identities must be open and transparent;
    \item \textbf{Persistence:} User identities and claims must be long-lived;
    \item \textbf{Portability:} Information and services about identities must be transportable;
    \item \textbf{Interoperability:} User identities and claims should be usable in different niches, organizations, and even countries;
    \item \textbf{Consent}: A user must agree to the use and disclosure of their identities and claims;
    \item \textbf{Minimalization:} Disclosure of identities and claims must be minimized to support privacy;
    \item \textbf{Protection:} The rights of the user must be protected
\end{enumerate}

\subsection{Related work}

In recent years, there have been several authors concerned with issues related to medical data sharing. Azaria \textit{et al.}~\cite{Azaria_2016} was the first to publish a fully functional prototype of a Blockchain system, called ``MedRec'', to manage medical records. Yue \textit{et al.} (2016) devised a similar solution that acts as a gateway between the health entities and the medical records by enabling patients to be owners of these records, allowing them to grant access privileges to desired hospitals and physicians. Both Azaria \textit{et al.} (2016) and Yue \textit{et al.} (2016) regarded patient ownership of their health information as a paramount factor in their works. Conceicao \textit{et al.}~\cite{Conceicao_2018} designed a general architecture for a global-scale EHR system based on Blockchain. Harding and Kotz~\cite{Hardin_2019} set out a number of essential research questions regarding EHRs and Blockchain that they could not find answers to, in their survey.

Kassab \textit{et al.}~\cite{kassab_2019} noted the challenges of adopting Blockchain as the core technology for EHR systems after carrying out a systematic literature review on this issue. This raised challenges like scalability and performance issues, as well as problems over usability, secure identification, and lack of incentives. 

Houtan \textit{et al.}~\cite{Houtan_2020} surveyed the state-of-the-art in a search for self-sovereign factors in patient data management solutions based on Blockchains. They concluded that most implementations looked for a hybrid model of distributed (DLTs like Blockchain) and centralized components to handle the trade-offs between patient privacy and medical record sharing capabilities.

Hussien \textit{et al.}~\cite{Hussien_2019} conducted a review of works that use Blockchain in the Health domain. The work examines how the healthcare applications are integrated with a Blockchain from several perspectives, ranging from the standpoint of the high-level architecture view to the low-level protocols and algorithms. In addition, the work provides a flowchart that determines if the Blockchain can be applied to a healthcare system. The work explores several technical problems, and offers possible solutions related to the security, privacy, integrity, access control, and interoperability by using Blockchain in the Health domain.

Aguiar \textit{et al.}~\cite{Aguiar_2019} carried out a survey of works that apply Blockchain in the area of healthcare. Their analysis focuses on the SWOT Table (Strength, Weakness, Opportunities, and Threats) of using the Blockchain in order to conduct research on the key points highlighted in this table. The work complements the other studies listed above because it describes how the Internet of Things (IoT) can be applied to various healthcare deployments for monitoring patients. It also shows how privacy and access control can be integrated with the Blockchain to allow users to access the electronic health records. With regard to this, the work examines several technologies, such as zero-knowledge proof~\cite{Goldreich_1994}, attributed-based encryption~\cite{Goyal_2006}, and standards, such as the HIPAA~\cite{hipaa_2020}, to protect the privacy data. 

Hasselgren \textit{et al.}~\cite{Hasselgren_2020} provide a review of works in which a Blockchain could improve the health domain operations (processes and services). The study explores the challenges raised when the data is shared among these operations and by several health sectors. The main factors studied were data access control, interoperability, provenance, and integrity. The Blockchain is used to overcome all of these problems in different areas, such as patient monitoring, storing medical record, and assisted decision-making, among others factors.

To the best of the authors' knowledge, no other reviews or surveys approached the health information storage and sharing challenges using self-sovereign identity models over Blockchain.

\section{Research Methodology}
\label{sec:method}

This section describes a protocol for systematic literature review that was carried out in this study.

It follows the guidelines set out by Kitchenham and Charters~\cite{Kitchenham_2007} for conducting systematic literature reviews (SLR). The SLR methodology comprises of three stages:
\begin{itemize}
    \item planning the review
    \item conducting the review
    \item reporting the review
\end{itemize}

\begin{figure}
    \centering
    \includegraphics[width=0.48\textwidth]{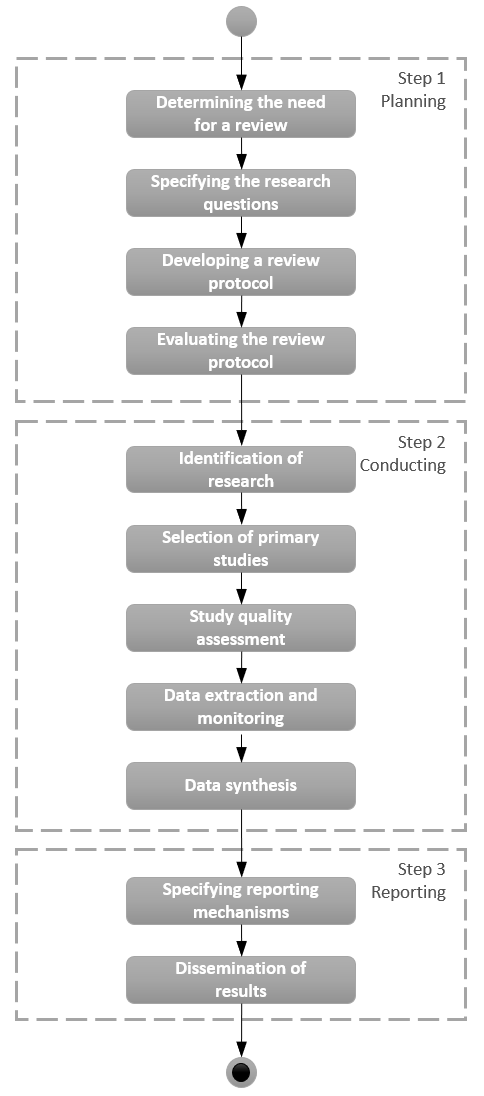}
    \caption{The stages of the systematic review, adapted from Kitchenham and Charters~\cite{Kitchenham_2007} }
    \label{fig:slrprotocol}
\end{figure}

Figure \ref{fig:slrprotocol} illustrates the interactions and the details of the stages followed by this study. 

The need for an SLR is supported by the fact that it involves research topics that may not be fully covered by current studies: understanding the research gaps and measures taken by academic research, can help to narrow the scope of the subject and raising key research questions, in particular with regard to this SLR, such as the following:

\begin{itemize}
    \item \textbf{RQ1:} What is the present research and implementations approaches of electronic health record data that employ Blockchain technology as the core mechanism for health information storage?
    \item \textbf{RQ2:} What types of information are currently being stored in the blocks of the Blockchain?
        \begin{itemize}
            \item \textbf{RQ2.1:} Are the health record metadata stored in the Blockchain in a searchable way?
        \end{itemize}
    \item \textbf{RQ3:} What mechanisms do the studies recommend as a means of ensuring the privacy of electronic health records?
        \begin{itemize}
            \item \textbf{RQ3.1:} Are there any studies that adopt Self-Sovereign Identity (SSI) strategies to handle electronic health records?
        \end{itemize}
    \item \textbf{RQ4:} How does the current research regard the data ownership and access control aspects of sharing electronic health records?
        \begin{itemize}
            \item \textbf{RQ4.1:} Who can write health records to the Blockchain?
            \item \textbf{RQ4.2:} Who can read the health records obtained from the Blockchain?
        \end{itemize}
\end{itemize}

The benefits of using Blockchains to handle the storage of health information are described by numerous papers, mainly because of its decentralized, immutable and traceable nature. RQ1 raises the question of how to distinguish between the strategies and technologies currently being used to store and maintain medical information in a Blockchain. The purpose of questions RQ2 and RQ2.1 is to help understand how health data is stored in the Blockchain and find out what search capabilities are available. It also helps to work out ways to handle large chunks of data, such as the results of diagnostic imaging files, in lightweight data structures like Blockchain. We recommend RQ3 to determine which privacy protection mechanisms have been employed to deal with sensitive information such as patient data. RQ3.1 seeks to determine self-sovereign identity features employed by selected studies. In addition, the aim of RQ4 is to map out the situations where Blockchain is used to store and share electronic health records and identify which parties are involved. RQ4.1 and RQ4.2 focus on some aspects of self-sovereign principles, specifically to the ownership of the data and the client consent procedures that arise from the current research.

A systematic review protocol was devised from the described research questions, in accordance with the  PICOC criteria (Population, Intervention, Comparison, Outcomes, Context) for the selection of effective primary studies, as recommended by Kitchenham and Charters~\cite{Kitchenham_2007} and Petticrew~\cite{Petticrew_2008}:
\begin{itemize}
    \item \emph{Population}: peer-review articles that describe Blockchain-based techniques for storing and sharing health data.
    \item \emph{Intervention}: identify the technologies and strategies adopted to handle sensitive medical information
    \item \emph{Comparison}: it does not apply
    \item \emph{Outcomes}: comprehensive interoperability while respecting the patient's privacy
    \item \emph{Context}: personal health records - medical appointments, medical diagnosis/treatment, immunization tracking, biomedical information collected by wearable devices, historical records of medical prescription, and any other background information that could give a picture of the patient's digital health.
\end{itemize}

The online tool Parsifal~\cite{Parsifal_2020} was adopted to define the protocol and to support the SLR.

\subsection{The search strategy}

The adopted search strategy involved carrying out automatic searches conducted through the search engines of four digital libraries: ACM Digital Library, IEEE Xplore, ScienceDirect, and SpringerLink. The following search terms were used to select the initial studies:

\hfill

\emph{``electronic health records'' AND (``self-sovereign identity'' OR blockchain) AND privacy AND interoperability}

\hfill

The search terms derived from the research questions were described earlier with the aim of selecting relevant primary studies. The term "blockchain" was combined with the OR operator to increase the number of selected studies since the usage of the "self-sovereign identity" phrase alone brought no results. Also, the term ``interoperability'' replaced the original term ``sharing'' after  the results of the first trials had been examined: the latter word seemed to be less effective for filtering unrelated studies (e.g., search noise). In addition, since it is a new subject, the period was restricted to studies published between 2017 and 2020. Table \ref{table:selectedstudies} summarizes the captured studies in the searches conducted on Jun 06th, 2020.

\begin{table}
\renewcommand{\arraystretch}{1.2}
\caption{Captured Studies by Source}
\label{table:selectedstudies}
\centering
\begin{tabular}{ p{4cm} | r }
 \hline
 \textbf{Source} & \textbf{Captured studies} \\ 
 \hline
 ACM Digital Library & 15 \\  
 IEEE Xplore & 110 \\
 ScienceDirect & 84 \\  
 SpringerLink & 114 \\ 
 \hline
 \textbf{Total} & \textbf{323} \\
 \hline
\end{tabular}
\end{table}

\subsection{Study selection}

The SLR protocol adopts inclusion and exclusion criteria to narrow down the number of studies submitted to a full-text analysis. The inclusion criteria are essential to restrict the scope of the selected studies, and this should help to answer the SLR research questions. The exclusion criteria allow unwanted studies to be removed, by listing the undesired or irrelevant types, languages, size, and subjects. A comprehensive list of the inclusion and exclusion criteria adopted by this review is displayed in Table \ref{table:selectioncriteria}.

\begin{table}
\renewcommand{\arraystretch}{1.2}
\caption{Inclusion / Exclusion Criteria for Study Selection}
\label{table:selectioncriteria}
\centering
\begin{tabular}{ r | p{12.6cm} }
 \hline
 \textbf{\#} & \textbf{Inclusion criteria} \\ 
 \hline
 1 & Studies that describe an architecture that uses Blockchain as the core mechanism for the storage of electronic health records \\  
 2 & Studies that are currently implementing software for managing electronic health records \\
 3 & Studies that put forward a system where the patient is the owner of his/her medical information \\  
 \hline
 \hline
 \textbf{\#} & \textbf{Exclusion criteria} \\ 
 \hline
 1 & Non-primary studies \\  
 2 & Filter noise (indexes / catalog articles)\\
 3 & Inaccessible studies \\  
 4 & Short studies (less than five pages) \\
 5 & Non-English written studies \\
 6 & Studies that do not consider patient privacy in the outlined design \\
 7 & Studies that do not include a serious discussion about how Blockchain technology can act as a mechanism for the storage of electronic health records \\
 8 & Studies that do not adopt an approach that involves medical record sharing between the patient and at least one of the following entities: doctors/physicians, hospitals, diagnostic lab centers, health insurance companies and immunization centers. \\
\hline
\end{tabular}
\end{table}

The 323 studies captured from the digital libraries search were screened for duplicates, short papers, book chapters, and inaccessible articles, and 140 were excluded. Of the 183 that remained, 98 studies were excluded after applying the inclusion/exclusion criteria to an analysis of the titles/abstracts. The remaining 85 papers were selected for a full-text review and analysis.

\subsection{Quality assessment}

Assessing the quality of primary studies is a crucial stage of the SLR process. It helps in the following ways: a) to improve the inclusion/exclusion criteria, b) to investigate the influence of quality on the findings, c) to sort out the selected studies in order of importance, d) to guive guidance on how the findings should be interpreted, and e) to make recommendations for further research~\cite{Kitchenham_2007}. This SLR ensured the quality of the selected papers by adapting the methods described by Ivarsson and Gorschek~\cite{Ivarsson_2011} to evaluate the 87 chosen studies in terms of their rigor and relevance, by submitting them to the questionnaire presented in Table \ref{table:qualityassessment} during the full text review.

\begin{table}
\renewcommand{\arraystretch}{1.5}
\caption{Quality assessment questions for evaluating rigor and relevance of studies, adapted from Ivarsson and Gorschek~\cite{Ivarsson_2011} }
\label{table:qualityassessment}
\centering
\begin{tabular}{ r | p{10.4cm} | p{1.5cm} }
 \hline
 \textbf{\#} & \textbf{Question} & \textbf{Criterion} \\ 
 \hline
 QA01 & Does it compares its results with the results of other studies? & Rigor \\ 
 QA02 & Does it explains the context in which the research was conducted? & Rigor \\ 
 QA03 & Does it presents the discussion about its findings? & Rigor \\ 
 QA04 & Does it describes the employed research methodology? & Rigor \\ 
 QA05 & Does it clearly presents its objectives? & Rigor \\ 
 QA06 & Does it considers the access control mechanisms and the security risks regarding unauthorized medical information disclosure? & Relevance \\ 
 QA07 & Does it raises concerns about patient health information privacy? & Relevance \\ 
 QA08 & Does it considers patient-generated health records? & Relevance \\ 
 QA09 & Does it considers the usage of exposed anonymized medical data for medical research purposes? & Relevance \\ 
 QA10 & Does it reviews or employ concepts of self-sovereign identities for electronic medical records storage? & Relevance \\ 
 QA11 & Does it considers the extrapolation of its objects to other industries? & Relevance \\ 
 QA12 & Does it raises any innovative approach for electronic medical records storage? &  Relevance \\ 
\hline
\end{tabular}
\end{table}

All the questions provided multiple-choice answers (``Yes'', ``Partially'', and ``No''), and these answers were scored as 1, 0.5 and 0 marks, respectively. The final selection only accepted those studies that scored 7.0 marks (67\%) or higher: 30 papers in total. The selected papers, their respective scores and citation count is shown in Table \ref{table:listofselectpapersqa} in Appendix A.

\begin{figure}
    \centering
    \includegraphics[width=1\textwidth]{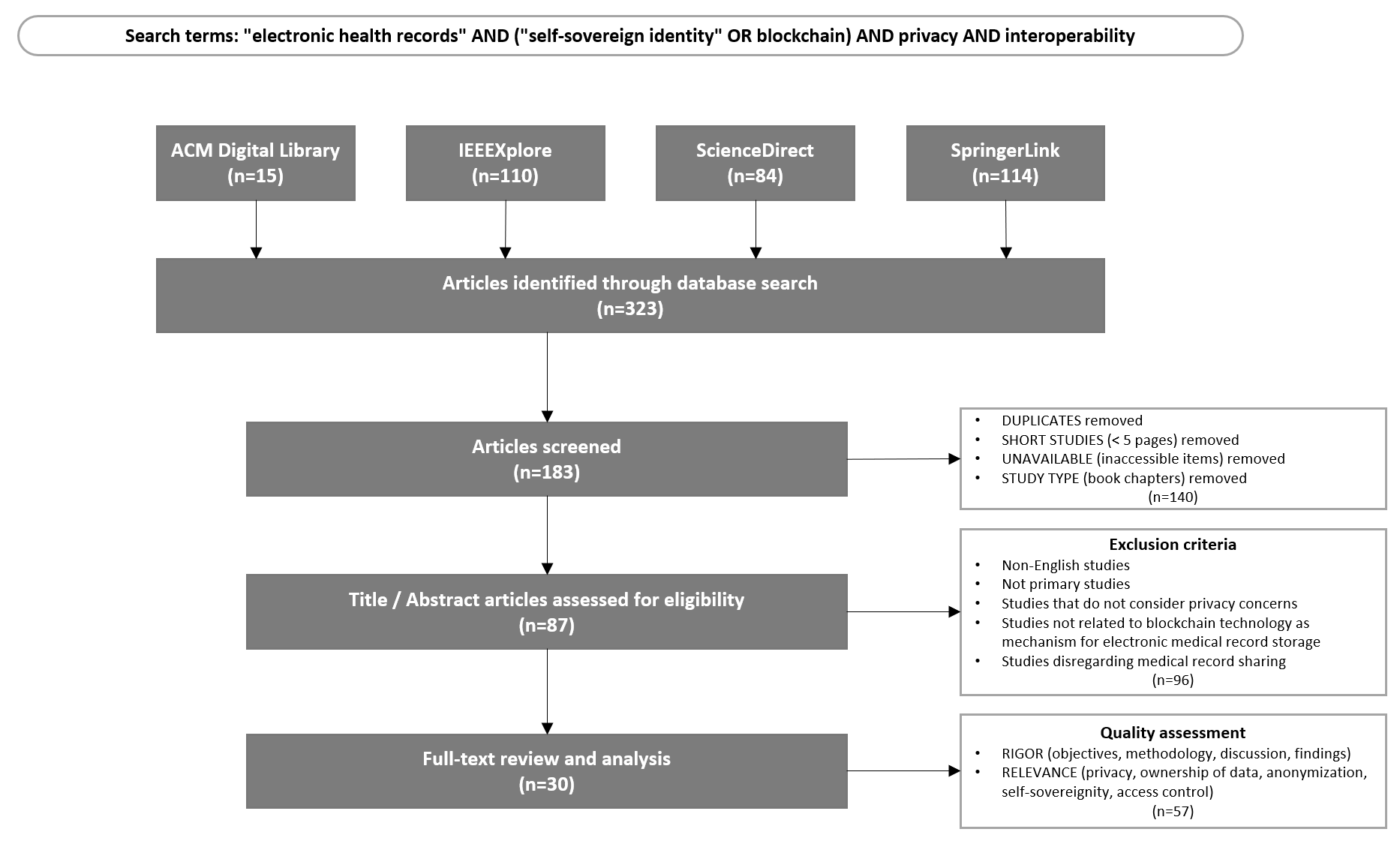}
    \caption{Study selection criteria}
    \label{fig:selectioncriteria}
\end{figure}

\subsection{Data extraction}

The SLR protocol included a data extraction form to assist in the unbiased collection of information from each of the 30 primary studies selected in this review. The information is divided into categories which are designed to answer the research questions shown in this systematic review. Table \ref{table:extractionform} outlines the data extraction form.

\begin{table}[t]
\renewcommand{\arraystretch}{1.2}
\caption{Data extraction form }
\label{table:extractionform}
\centering
\begin{tabular}{ r | p{4cm} | p{6cm} | p{2.2cm}}
 \hline
 \textbf{\#} & \textbf{Information Category} & Description & Related RQ \\ 
 \hline
 1 & Study Id & Unique identifier for the study & Study overview\\ 
 2 & Type of article & In proceedings article, journal article & Study overview \\ 
 3 & Title & Metadata & Study overview \\ 
 4 & Authors & Metadata & Study overview \\ 
 5 & Publication year & Metadata & Study overview \\ 
 6 & Article source & ACM, IEEE Xplore, SpringerLink, ScienceDirect &  Study overview \\
 7 & Research type & Evaluation, opinion, prototype, solution proposal, validation research & Study overview \\
 8 & Context & EHR, PHR, patient monitoring, insurance payments, etc. & Study overview \\
 9 & Technology & Hyperledger, Ethereum, Bitcoin, custom & RQ1 \\
 10 & Content stored on-chain & Full content, metadata, pointer, hash & RQ2 / RQ2.1 \\
 11 & Content stored off-chain & None, full content, sidechain & RQ2 \\
 12 & Blockchain type & Public, permissioned, private & RQ1 / RQ3 \\
 13 & Paradigm & Transactional, Self-sovereign Identity, Sidechain & RQ3 \\
 14 & Access control & Claim, attribute-based, PKI, smart contract, ACL, others & RQ3 \\
 15 & Entities that generate health records & Patient, physician, hospital, diagnostic labs, health insurance & RQ4 / RQ4.1 \\
 16 & Entities with which health records are shared & Patient, physician, hospital, diagnostic labs, health insurance & RQ4 / RQ4.2 \\
 17 & Entities that own the health records & Patient, physician, hospital, diagnostic labs, health insurance & RQ4 \\
 18 & Entities with which anonymized health records are available & Public, patient, physician, hospital, diagnostic labs, health insurance & RQ4 / RQ4.2 \\
 \hline
\end{tabular}
\end{table}

\subsection {Threats to validity}

The guidelines for carrying out systematic literature reviews laid down by Kitchenham and Charters~\cite{Kitchenham_2007} enabled the authors to draw up a review protocol that mitigates the risk of biased results. However, the authors must assume that some aspects of the protocol may still threaten the validity of this study's conclusions.

In the selection of primary studies, the risk of failing to take account of papers was mitigated by refining the search string by conducting pilot searches in the selected digital libraries until the collected sample of papers showed a close relation to the research topics. The digital library selection was also taken into account: all four libraries are among the most popular scientific databases in the computer science and software engineering industry. Despite these attempts, an identified threat had to be considered: some papers had to be rejected during the selection phase because their content was not publicly available. 

\section{Results and Analysis}
\label{sec:results}

\begin{figure}
    \centering
    \includegraphics[width=0.6\textwidth]{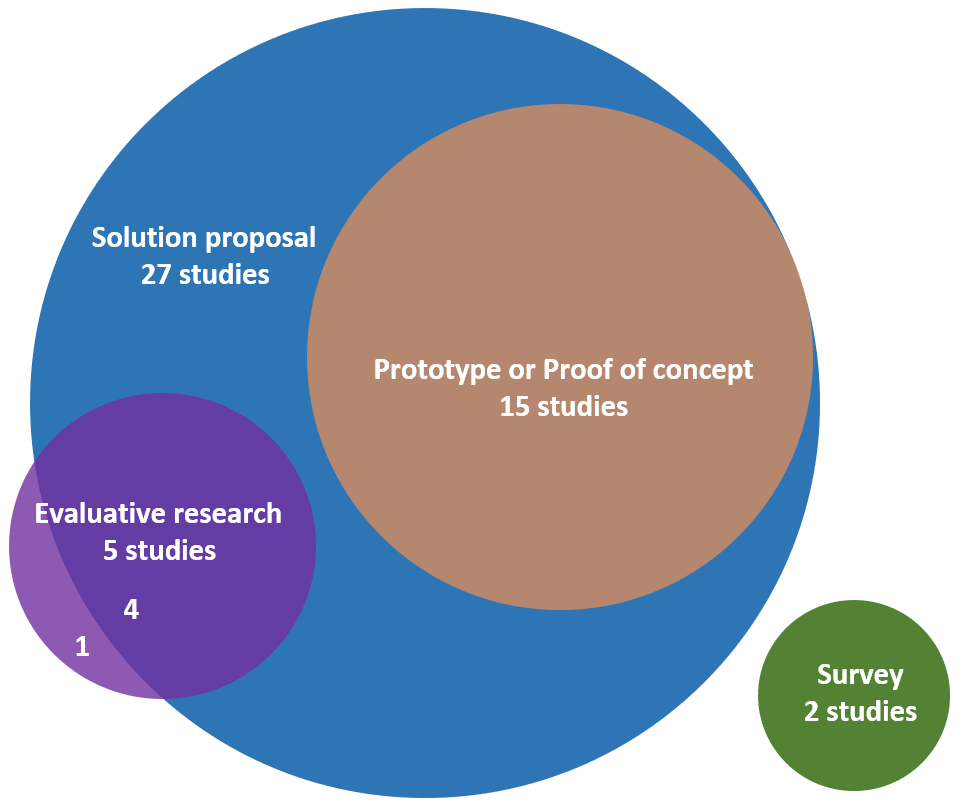}
    \caption{Study types}
    \label{fig:studydistribution}
\end{figure}

This section examines the results of this SLR and analyzes its findings, initially by providing an overview of the studies and then assessing the findings for each proposed research question. A total of 30 studies, related to the way Blockchain can be applied to storage and sharing of electronic health records, were selected for this analysis. Most of these studies comprise suggestions for solutions (27 papers), evaluative research (5 papers), and surveys (2 papers). Figure \ref{fig:studydistribution} illustrates the distribution of these papers in a Venn diagram on the basis of their categories.

Most of the proposal papers (56\%) displayed current solutions that are implemented as prototypes and proof of concepts. Those implementations showed the usefulness of the architecture and technology employed for these solutions. The evaluative papers made it possible to understand the key features and limitations of Blockchain technology when applied to healthcare data, while the survey studies offered a holistic approach to the state-of-the-art research in the field.

\begin{figure}[t]
    \centering
    \includegraphics[width=0.7\textwidth]{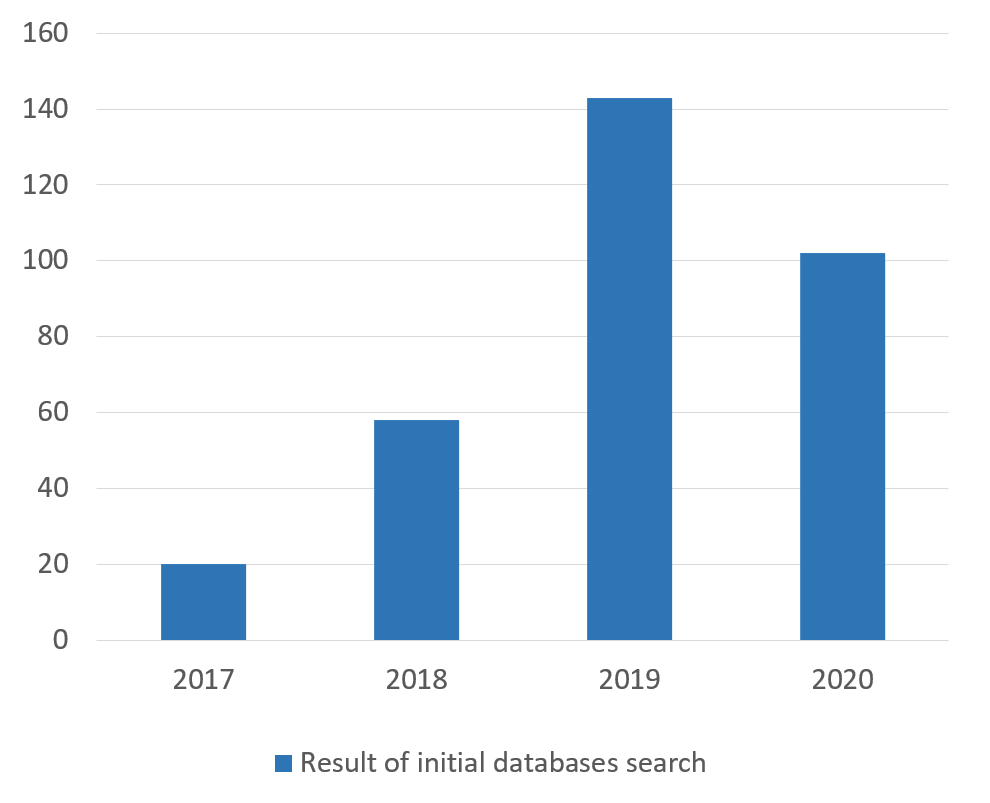}
    \caption{Distribution of studies over the years (left axis scale for results of initial search, right axis scale for final selection of studies}
    \label{fig:papersovertheyears}
\end{figure}

The increase in the number of published studies over the years reflects the novelty of the subject. The Figure \ref{fig:papersovertheyears} depicts the distribution of published studies over time. The period of time allocated for the search in the digital libraries (2017-2020)  and the fact that the articles of 2020 only represent papers published until mid-2020 (the period in which this study was carried out), shows that there is a growing interest in the subject.

In compiling the application context of the selected papers, this SLR devised a list of capabilities that comprise the classification of these papers, as shown in Table \ref{table:capabilities}. These capabilities are not mutually exclusive and describe the functional outcomes that the studies sought to achieve.

\begin{table}[ht]
\renewcommand{\arraystretch}{1.2}
\caption{List of selected papers and the capabilities they support.}
\label{table:capabilities}
\centering
\begin{tabular}{ p{4.8cm} | c | c | c | c | c | c | c }
 \hline
 Study & EHR & PHR & RES & MON & INS & TRD & PRC \\ 
 \hline
Zhuang \textit{et al.}~\cite{Zhuang_2020} &  & \cellcolor[HTML]{CCCCCC} &  &  &  &  & \\
 \hline
Houtan \textit{et al.}~\cite{Houtan_2020} & \cellcolor[HTML]{CCCCCC} & \cellcolor[HTML]{CCCCCC} &  &  &  &  & \\
 \hline
Roehrs \textit{et al.}~\cite{Roehrs_2019} & \cellcolor[HTML]{CCCCCC} & \cellcolor[HTML]{CCCCCC} &  &  &  &  & \\
 \hline
Dagher \textit{et al.}~\cite{Dagher_2018} & \cellcolor[HTML]{CCCCCC} &  &  &  &  &  & \cellcolor[HTML]{CCCCCC}\\
 \hline
Bhattacharya \textit{et al.}~\cite{Bhattacharya_2020} & \cellcolor[HTML]{CCCCCC} & \cellcolor[HTML]{CCCCCC} & \cellcolor[HTML]{CCCCCC} &  &  &  & \cellcolor[HTML]{CCCCCC}\\
 \hline
Azbeg \textit{et al.}~\cite{Azbeg_2018} &  & \cellcolor[HTML]{CCCCCC} &  & \cellcolor[HTML]{CCCCCC} &  &  & \\
 \hline
Hardin and Kotz~\cite{Hardin_2019} & \cellcolor[HTML]{CCCCCC} & \cellcolor[HTML]{CCCCCC} &  &  &  &  & \\
 \hline
Uddin \textit{et al.}~\cite{Uddin_2020} &  & \cellcolor[HTML]{CCCCCC} &  & \cellcolor[HTML]{CCCCCC} &  &  & \\
 \hline
Gordon and Catalini~\cite{Gordon_2018} & \cellcolor[HTML]{CCCCCC} & \cellcolor[HTML]{CCCCCC} &  &  &  &  & \\
 \hline
Sharma \textit{et al.}~\cite{Sharma_2020} & \cellcolor[HTML]{CCCCCC} &  &  &  & \cellcolor[HTML]{CCCCCC} &  & \\
 \hline
Wang \textit{et al.}~\cite{Wang_2019} & \cellcolor[HTML]{CCCCCC} & \cellcolor[HTML]{CCCCCC} &  &  &  &  & \\
 \hline
Zhou \textit{et al.}~\cite{Zhou_2018} &  &  & \cellcolor[HTML]{CCCCCC} &  &  & \cellcolor[HTML]{CCCCCC} & \\
 \hline
Li \textit{et al.}~\cite{Li_2019} & \cellcolor[HTML]{CCCCCC} &  & \cellcolor[HTML]{CCCCCC} &  &  & \cellcolor[HTML]{CCCCCC} & \\
 \hline
Xiao \textit{et al.}~\cite{Xiao_2018} & \cellcolor[HTML]{CCCCCC} &  & \cellcolor[HTML]{CCCCCC} &  &  &  & \\
 \hline
Griggs \textit{et al.}~\cite{Griggs_2018} & \cellcolor[HTML]{CCCCCC} &  &  & \cellcolor[HTML]{CCCCCC} &  &  & \\
 \hline
Mubarakali~\cite{Mubarakali_2020} & \cellcolor[HTML]{CCCCCC} &  &  & \cellcolor[HTML]{CCCCCC} & \cellcolor[HTML]{CCCCCC} &  & \\
 \hline
Marangappanavar and Kiran~\cite{Marangappanavar_2020} &  & \cellcolor[HTML]{CCCCCC} &  &  &  &  & \\
 \hline
Huang \textit{et al.}~\cite{Huang_2019} & \cellcolor[HTML]{CCCCCC} &  &  &  &  &  & \\
 \hline
Pournaghi \textit{et al.}~\cite{Pournaghi_2020} & \cellcolor[HTML]{CCCCCC} & \cellcolor[HTML]{CCCCCC} &  &  &  &  & \\
 \hline
Roehrs \textit{et al.}~\cite{Roehrs_2017} & \cellcolor[HTML]{CCCCCC} & \cellcolor[HTML]{CCCCCC} &  &  &  &  & \\
 \hline
Buzachis \textit{et al.}~\cite{Buzachis_2019} & \cellcolor[HTML]{CCCCCC} & \cellcolor[HTML]{CCCCCC} &  &  &  &  & \\
 \hline
Toshniwal \textit{et al.}~\cite{Toshniwal_2019} & \cellcolor[HTML]{CCCCCC} &  &  &  &  &  & \\
 \hline
Talukder \textit{et al.}~\cite{Talukder_2018} & \cellcolor[HTML]{CCCCCC} &  & \cellcolor[HTML]{CCCCCC} &  &  &  & \\
 \hline
Tripathi \textit{et al.}~\cite{Tripathi_2020} & \cellcolor[HTML]{CCCCCC} & \cellcolor[HTML]{CCCCCC} & \cellcolor[HTML]{CCCCCC} & \cellcolor[HTML]{CCCCCC} &  &  & \\
 \hline
Donawa \textit{et al.}~\cite{Donawa_2019} & \cellcolor[HTML]{CCCCCC} &  &  &  &  &  & \\
 \hline
Mahore \textit{et al.}~\cite{Mahore_2019} & \cellcolor[HTML]{CCCCCC} &  & \cellcolor[HTML]{CCCCCC} &  &  &  & \\
 \hline
Guo \textit{et al.}~\cite{Guo_2018} & \cellcolor[HTML]{CCCCCC} &  &  &  &  &  & \\
 \hline
Jin \textit{et al.}~\cite{Jin_2019} & \cellcolor[HTML]{CCCCCC} &  & \cellcolor[HTML]{CCCCCC} &  &  &  & \\
 \hline
Zhang and Lin~\cite{Zhang_2018} & \cellcolor[HTML]{CCCCCC} &  &  &  &  &  & \\
 \hline
Shahnaz \textit{et al.}~\cite{Shahnaz_2019} & \cellcolor[HTML]{CCCCCC} &  &  &  &  &  & \\
\hline
 \multirow{2}{0.8cm}{Studies} & 25 & 14 & 8 & 5 & 2 & 2 & 2 \\
 & 83.3\% & 46.7\% & 26.7\% & 16.7\% & 6.7\% & 6.7\% & 6.7\% \\
 \hline
\end{tabular}
\end{table}

\textbf{Electronic health records (EHR)}: this is a compilation of studies that entail improving the management of digital records and sharing processes among healthcare entities. Most articles (83.3\%) fell into this category, since they describe a viable means of balancing the need to create a decentralized platform to exchange the medical records of patients while keeping sensitive information private and secure. These studies recognized the features and benefits of the Blockchain technology to accomplish their goals.

\textbf{Personal health records (PHR)}: this is a compilation of studies designed to secure the ownership of health records for the patients, not only by allowing them to produce health data themselves, but also by granting them the prerogative to decide who can access their data and when. 46.7\% of the articles describe Blockchain-based techniques that put patients at the center of their data, but only two articles (6.7\%) apply concepts of self-sovereign identity to achieve this outcome.

\textbf{Medical research/Public health (RES)}: this capability includes studies concerning the availability of anonymous public medical records for scientific research and the planning of public health policies. Eight studies (26.7\%) address this matter. 

\textbf{Patient monitoring (MON)}: this capability includes articles that seek decentralized methods for collecting medical information from patients in special conditions - like those in intensive care units or patients with chronic diseases that must constantly monitor their condition. From the selected studies, 16.7\% (5 studies) fall into this category.

\textbf{Insurance payments (INS)}: this sector is investigated by two studies (6.7\%) that seek mechanisms for automating health insurance claims and payments, as well as making use of the traceability features of Blockchain-based technologies to prevent fraud.

\textbf{Healthcare data trading (TRD)}: this is a system that studies how patients can reaffirm ownership of their health data. This can be carried out in a marketplace where interested parties can negotiate financial deals in exchange for private medical information. Public and permissioned Blockchains support these negotiations that range from medical research incentives to the offer of discounts in health insurance charges.

\textbf{Control methods for prescription drugs (PRC)}: this area includes studies that consider restrictions on the prescription of controlled drugs. Two studies (6.7\%) examine the immutability features of Blockchain to address this question.

%RQ1
\subsection{RQ1: What is the present research and implementations approaches of electronic health record data that employ Blockchain technology as the core mechanism for health information storage?}

As mentioned, Blockchain has properties that allow following some SSI principles, such as transparency and persistence. The aim of this research question is to determine a) which of the selected studies provides a design solution that is mature enough to identify the key technologies from the Blockchain ecosystem and b) how these technologies have been employed to face the challenges of storing and sharing electronic health records in a Blockchain.

\subsubsection{Results}

\begin{figure}[h]
    \centering
    \includegraphics[width=0.75\textwidth]{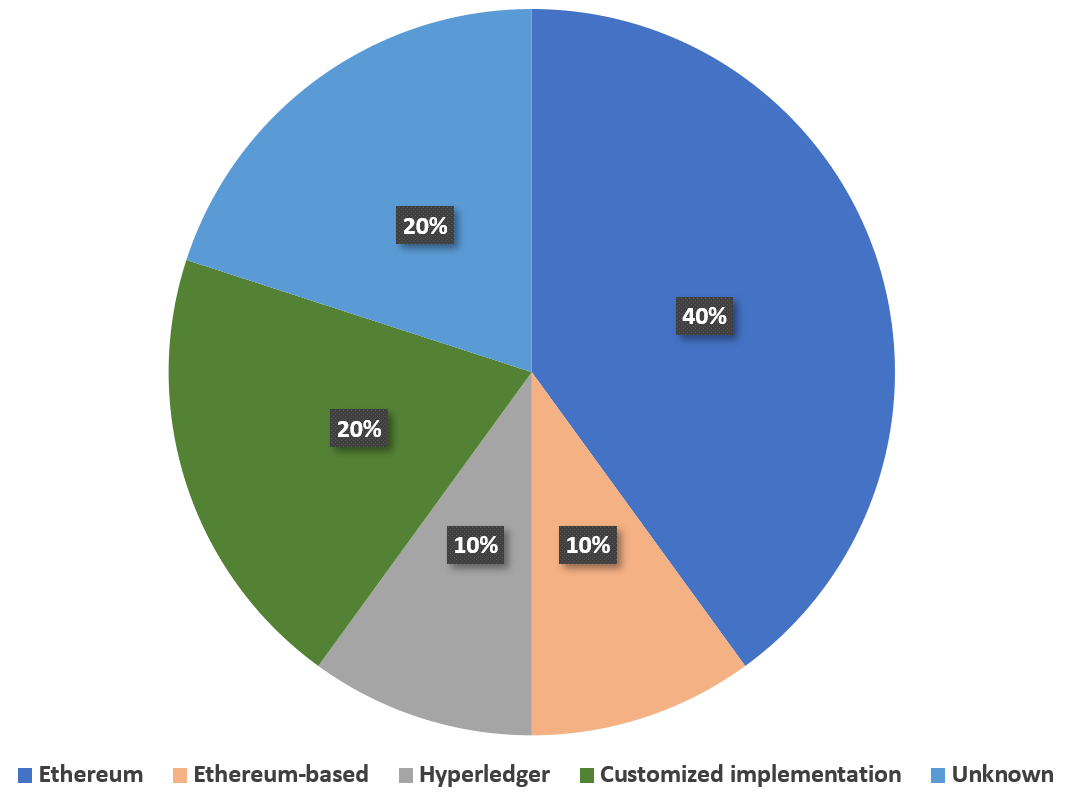}
    \caption{Blockchain technologies employed by proposal and prototype studies}
    \label{fig:technologies}
\end{figure}

Most of the selected studies (27) are, in some way, currently putting forward schemes to address the question of how to manage the storage and exchange of electronic health records. However, half of these studies (15) outlined prototypes and proof of concepts in their works. Figure \ref{fig:technologies} displays the range of technologies that they used.

Blockchain-based frameworks underpinned most of the solution designs, particularly those that provided smart contracts support: Ethereum and Ethereum-based platforms were selected as core elements of 15 studies~\cite{Zhuang_2020}, \cite{Dagher_2018}, \cite{Azbeg_2018}, \cite{Sharma_2020}, \cite{Wang_2019}, \cite{Zhou_2018}, \cite{Griggs_2018}, \cite{Mubarakali_2020}, \cite{Marangappanavar_2020}, \cite{Buzachis_2019}, \cite{Toshniwal_2019}, \cite{Talukder_2018}, \cite{Jin_2019}, \cite{Zhang_2018}, \cite{Shahnaz_2019} while Hyperledger frameworks featured as the central component in 3 studies~\cite{Xiao_2018}, \cite{Huang_2019}, \cite{Mahore_2019}. Specific blockchain technologies were used for prototype-based studies like Corda~\cite{Bhattacharya_2020}, OPNET \cite{Pournaghi_2020} and other customized implementations~\cite{Uddin_2020}. 

\subsubsection{Analysis and discussion}

The idea of introducing systems to digitalize patients' medical records is not new: healthcare service providers have been using electronic health record systems for some time but, since there is a lack of standardization, each of these repositories had turned into a silo of medical information. Huang \textit{et al.}~\cite{Huang_2019} describes how in 2016 the New Zealand Ministry of Health launched a program to create a nationwide smart health system to address the diversity of health applications: ``One of the core components of the programme is to create an EHR that collates a patient's health information into a single longitudinal `health story' accessible to all healthcare stakeholders.''~\cite{Huang_2019} The challenges of achieving this goal go beyond interoperability: Hardin and Kotz~\cite{Hardin_2019} explain how Blockchain-based technologies could help. These challenges include the following~\cite{Hardin_2019}:

\begin{itemize}
    \item Access control
    \item Data integrity
    \item Interoperability
    \item Auditing
\end{itemize}

Distributed Ledger Technologies (DLT) like Blockchain offer some features that can overcome these problems. They are \textbf{immutable}, as Hardin and Kotz recognize: ``the simple and obvious solution to maintaining integrity for an individual's health data is to store it on the blockchain''~\cite{Hardin_2019}. This feature allows Blockchain-based techniques to be applied in situations that require trust among peers even though there is no central authority to enforce it. Examples of these techniques are the prescription control situations described by Dagher \textit{et al.}~\cite{Dagher_2018} and Bhattacharya \textit{et al.}~\cite{Bhattacharya_2020} or insurance claims management like that out forward by Mubarakali~\cite{Mubarakali_2020} and Sharma \textit{et al.}~\cite{Sharma_2020}.

DLTs are also \textbf{decentralized}, which means that distinct entities can act as nodes in a network and share the same copy of the records. Nodes in a Blockchain network can add new entries to the ledger and receive updates to their local version of the ledger with integrity assurance. This feature supports use cases like the use of Internet of Medical Things (IoMT) devices for patient health monitoring or wearable devices for wellness information like those recommended by Azbeg \textit{et al.}~\cite{Azbeg_2018}, Uddin \textit{et al.} \cite{Uddin_2020}, Griggs \textit{et al.}~\cite{Griggs_2018}, Mubarakali~\cite{Mubarakali_2020}, and Tripathi \textit{et al.}~\cite{Tripathi_2020}

Blockchains arrange their information in connected blocks of data and, as each node in the network has a copy of all blocks, the Blockchain is also \textbf{auditable}. Studies like that of Toshniwal \textit{et al.}\cite{Toshniwal_2019} take advantage of this feature through a medical records exchange system that keeps track of the movement of electronic medical records between the healthcare providers.

Even with the search string used for the capture of studies that employed Blockchain technologies, some selected papers adopted other decentralized approaches: Roehrs \textit{et al.} \cite{Roehrs_2017, Roehrs_2019} employ a decentralized system based on Apache Kafka ecosystem products, while Li \textit{et al.}\cite{Li_2019} describe an Edge computing network of server nodes to form a decentralized design.

%RQ2
\subsection{RQ2: What types of information are currently being stored in the blocks of the Blockchain? \newline RQ2.1: Are the health record metadata stored in the Blockchain in a searchable way?}

As discussed in the previous section, Blockchain technologies offer many benefits but there are also some caveats. Owing to its decentralized nature, every node in the network has the same information copy, and thus large blocks of data could increase the requirements for space, bandwidth, and computational power~\cite{Marangappanavar_2020}, which makes Blockchain difficult to scale. The purpose of RQ2 is to find out how the selected studies determine what information goes into the Blockchain blocks and what is stored elsewhere. RQ2.1 seeks to identify the metadata structures and standards employed by the selected studies to allow the records to be searched.

\subsubsection{Results}

Storing large amounts of information into the Blockchain is problematic, and this is particularly true with regard to health data because of the size of the records. Laboratory exams, for example, usually produce large image files as their results. Thus, it is a convention to specify which types of health information are \textit{on-chain} (i.e., inside a block of the chain) and which are \textit{off-chain} (i.e., in an external, generally distributed, storage system).

\begin{figure}[ht]
    \centering
    \includegraphics[width=0.7\textwidth]{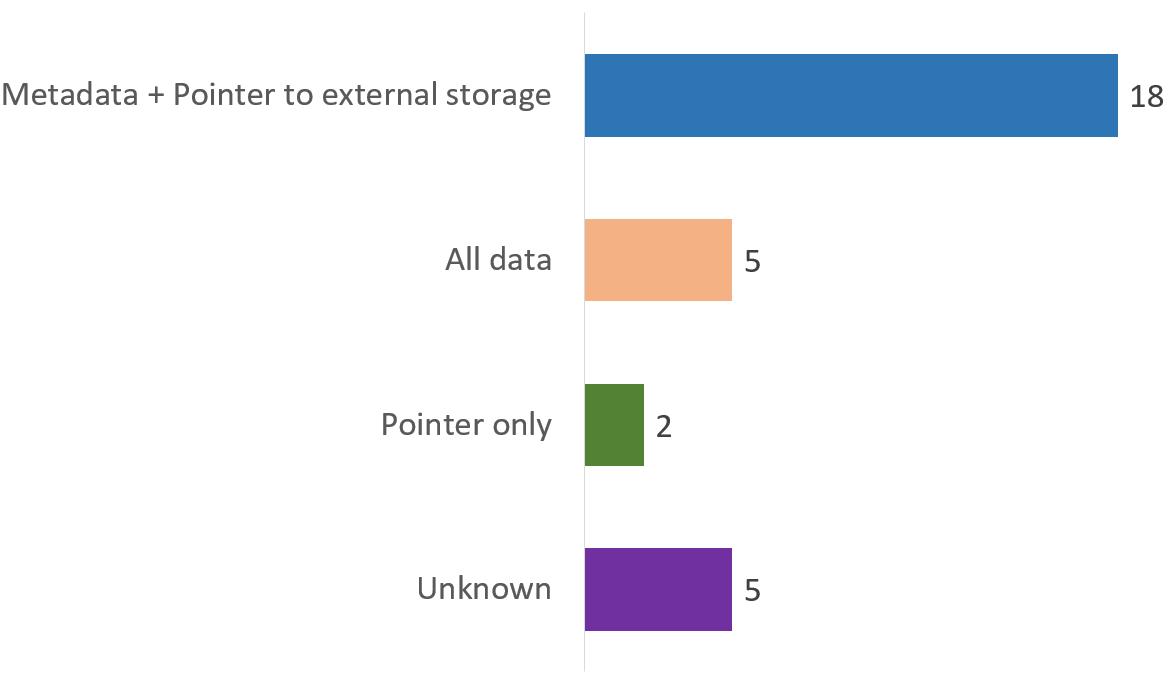}
    \caption{Distribution pattern of the schemes for on-chain information storage}
    \label{fig:onchainoffchain}
\end{figure}

Most selected studies (19) recommend using the on-chain transactions to store metadata about the health records, the hash value of the original record, and a pointer to an off-chain storage mechanism where the actual record is kept \cite{Zhuang_2020, Dagher_2018, Uddin_2020, Sharma_2020, Zhou_2018, Xiao_2018, Marangappanavar_2020, Pournaghi_2020, Buzachis_2019, Toshniwal_2019, Talukder_2018, Tripathi_2020, Donawa_2019, Mahore_2019, Jin_2019, Zhang_2018, Shahnaz_2019}. Two studies~\cite{Azbeg_2018, Griggs_2018} only keep a link to the off-chain record file. There are six studies, however, that embed all the health information in the on-chain transaction~\cite{Roehrs_2017, Roehrs_2019, Bhattacharya_2020, Wang_2019, Mubarakali_2020, Huang_2019}. Five studies are not clear about this subject. Figure \ref{fig:onchainoffchain} illustrates these approaches and the distribution pattern of the studies.

\subsubsection{Analysis and discussion}

The definition of where the information is kept determines the features that an electronic health records system can provide, and the studies selected by the SLR had to address this trade-off: keeping relevant data (metadata) in the Blockchain can lead to feature-rich solution that is hard to scale. 

Donawa \textit{et al.}~\cite{Donawa_2019} examine this matter in detail, by using the idea of side chains to address this trade-off. Side chains are auxiliary Blockchain structures bound to a main Blockchain that organize the information through the use of metadata. The argument is that a single Blockchain data structure would act as a bottleneck in large-scale scenarios, where multiple nodes of the network have to reach a consensus. Studies from Pournaghi \textit{et al.}~\cite{Pournaghi_2020}, Tripathi \textit{et al.}~\cite{Tripathi_2020}, and Zhang and Lin~\cite{Zhang_2018} also based their solutions on the side chain approach.

Keeping electronic health records metadata on-chain is convenient for the following reasons:

\begin{itemize}
    \item With the aid of the metadata parameters like record types, patient condition, demographics, and others, it can help locate and filter the health records. Zhou \textit{et al.}~\cite{Zhou_2018} find a solution that combines Blockchain and machine learning to create a marketplace of private healthcare records through indexing and grouping through the principle of similarity. Talukder \textit{et al.}~\cite{Talukder_2018} discuss a ``proof-of-disease'' protocol, using ``medical miners'' to evaluate artificial intelligence-produced diagnoses based on healthcare records metadata.
    \item It enables features powered by smart contracts like access control, data encryption/decryption, record integrity checking, and others. Jin \textit{et al.}~\cite{Jin_2019} designed an attribute-based encryption mechanism for sharing healthcare records using smart contracts. Pournaghi \textit{et al.}~\cite{Pournaghi_2020} and Marangappanavar \textit{et al.}~\cite{Marangappanavar_2020} used smart contracts to control access to sensitive healthcare information.
    \item It drives interoperability by employing health records standards like OpenEHR,  HL7 FHIR, and ISO EN13606. Houtan \textit{et al.} summarizes the question as follows~\cite{Houtan_2020}: ``Data can be interpreted by human operators but not by HIS (Health Information Systems). In structural standards, exchanged data fields can be interpreted; this means that  data types and fields are recognizable by all participating HIS.''
    \item Client applications only have to fetch large files for the records they need; they use metadata information to select the relevant entries and downloading raw data files from external sources, like Inter-Planetary File Systems structures (IPFS). The peer-to-peer capabilities of IPFS are used in the EHR solutions described by Azbeg \textit{et al.}~\cite{Azbeg_2018}, Sharma \textit{et al.}~\cite{Sharma_2020}, Marangappanavar and Kiran.~\cite{Marangappanavar_2020}, Buzachis \textit{et al.}~\cite{Buzachis_2019} and Shahnaz \textit{et al.}~\cite{Shahnaz_2019}
\end{itemize}

In the case of some studies, however, keeping the whole information on-chain is the preferred approach, as they have to address peculiar situations. Mubarakali and Kiran~\cite{Mubarakali_2020} set out a method for patient monitoring that only require biomedical scanners data to be collected and stored. Rhoers at al.~\cite{Roehrs_2017}~\cite{Roehrs_2019} designed a custom distributed structure (based on the Chord algorithm~\cite{Stoica_2003}) and claim that it has a better performance than traditional Blockchain approaches: their performance test shows that the solution is capable of scaling properly under a heavy load.

%RQ3
\subsection{RQ3: What mechanisms do the studies recommend as a means of ensuring the privacy of electronic health records? \newline RQ3.1: Are there any studies that adopt Self-Sovereign Identity (SSI) strategies to handle electronic health records?}

This research question seeks to map out how studies plan to protect the privacy of patients' health information by adopting an inherently open technology like Blockchain. 

\subsubsection{Results}

With regard to the privacy of personal health information, this SLR divided the selected studies into two distinct groups: Blockchain types and access control mechanisms.

Blockchain types reflect the general availability of the Blockchain network. In this regard, the studies can be categorized as:
\begin{itemize}
    \item \textbf{Private:} a central authority controls the access to the network. 43.3\% of the selected studies use private Blockchains;
    \item \textbf{Permissioned:} also called Consortium Blockchains, the access to the network is controlled by member entities. 33.3\% of the selected studies refer to permission Blockchains;
    \item \textbf{Public:} everyone has access to the Blockchain network. 10\% of the studies use public blockchains.
\end{itemize}

Additionally, 13.3\% of the selected studies do not specify which Blockchain type is employed.

Blockchain types include an initial level of access restriction. However, an additional protection layer must be considered: these access control mechanisms ensure that, for its general audience, only authorized entities can access specific records. Access control mechanisms are not specific to Blockchain types, so Table \ref{table:accessheatmap} shows a Heat Map to represent these relations.

%\begin{figure}[t]
%    \centering
%    \includegraphics[width=0.96\textwidth]{assets/accessheatmap.png}
%    \caption{Study distribution based on access control mechanisms and Blockchain type - Heat Map}
%    \label{fig:accessheatmap}
%\end{figure}

\begin{table}[t]
\small
\renewcommand{\arraystretch}{1.2}
\caption{Study distribution based on access control mechanisms and Blockchain type - Heat Map}
\label{table:accessheatmap}
\centering
\begin{tabular}{ | p{1.7cm} | p{2cm} | p{2cm} | p{2cm} | p{2.1cm} | p{2cm} | }
 \hline
 \multirow{2}{4em}{Blockchain type} & \multicolumn{5}{ c|}{Access control mechanism} \\
 \cline{2-6}
  & \makecell[c]{Smart\\Contracts} & \makecell[c]{Public Key\\Infrastructure} & \makecell[c]{Access\\Control List} & \makecell[c]{Attribute-based\\Encryption} & \makecell[c]{Unknown}  \\ 
 \hline
  Private & 
    \cellcolor[HTML]{000000}\textcolor{white}{\makecell[c]{ 7 }} & 
    \cellcolor[HTML]{BFBFBF}\makecell[c]{ 3 } & 
    \cellcolor[HTML]{BFBFBF}\makecell[c]{ 3 } & 
    \cellcolor[HTML]{FFFFFF}\makecell[c]{ 0 } & 
    \cellcolor[HTML]{FFFFFF}\makecell[c]{ 0 } \\
 \hline 
  Consortium & 
    \cellcolor[HTML]{404040}\textcolor{white}{\makecell[c]{ 6 }} & 
    \cellcolor[HTML]{FFFFFF}\makecell[c]{ 0 } & 
    \cellcolor[HTML]{F2F2F2}\makecell[c]{ 1 } & 
    \cellcolor[HTML]{D9D9D9}\makecell[c]{ 2 } & 
    \cellcolor[HTML]{F2F2F2}\makecell[c]{ 1 } \\
 \hline 
  Public & 
  \cellcolor[HTML]{D9D9D9}\makecell[c]{ 2 } & 
  \cellcolor[HTML]{F2F2F2}\makecell[c]{ 1 } & 
  \cellcolor[HTML]{FFFFFF}\makecell[c]{ 0 } & 
  \cellcolor[HTML]{FFFFFF}\makecell[c]{ 0 } & 
  \cellcolor[HTML]{FFFFFF}\makecell[c]{ 0 } \\
 \hline 
  Unknown & 
  \cellcolor[HTML]{FFFFFF}\makecell[c]{ 0 } & 
  \cellcolor[HTML]{FFFFFF}\makecell[c]{ 0 } & 
  \cellcolor[HTML]{FFFFFF}\makecell[c]{ 0 } & 
  \cellcolor[HTML]{F2F2F2}\makecell[c]{ 1 } & 
  \cellcolor[HTML]{BFBFBF}\makecell[c]{ 3 } \\
 \hline  
\end{tabular}
\end{table}

\subsubsection{Analysis and discussion}

Given the sensitivity of the information they are dealing with, most studies rely on private or permissioned Blockchains. National and international regulatory bodies like HIPAA and GDPR impose severe restrictions on Personally Identifiable Information (PII) and Protected Health Information (PHI).

Studies that adopt public Blockchains seek to comply with regulatory requirements: Tripathi \textit{et al.}~\cite{Tripathi_2020}, Talukder \textit{et al.}~\cite{Talukder_2018}, and Zhou \textit{et al.}~\cite{Zhou_2018} examine the question of using health data provided anonymously for medical research and clinical trials.

The most widely used and versatile instruments for controlling access to records in a Blockchain are smart contracts. Each of this is used as an intermediary component (called a proxy) to obtain decrypted information from encrypted block transactions, as described by Xiao \textit{et al.}~\cite{Xiao_2018} and Huang \textit{et al.}~\cite{Huang_2019}. Smart contracts are also used to determine whether a user was granted or denied access to another person's health records. Marangappanavar and Kiran~\cite{Marangappanavar_2020} devised an architecture which entailed using smart contracts to protect raw medical information stored in external IPFS repositories.

Sharma \textit{et al.}~\cite{Shahnaz_2019}, Wang \textit{et al.}~\cite{Wang_2019}, and Dagher \textit{et al.}~\cite{Dagher_2018} described the use of smart contracts in conjunction with proxy re-encryption technology to provide mechanisms to share health data with other parties while keeping sensitive information private. Sharma \textit{et al.}~\cite{Shahnaz_2019} describes proxy re-encryption as ``a type of public key encryption that allows an intermediary (proxy) to transform cipher text from one public key to another without being able to observe the plain data.''

Smart contracts are an inherent part of modern Blockchain technology and, thus, widely adopted. In addition to smart contracts, there are also Public Key Infrastructure (PKI) and Role-Based Access Control (RBAC) implementations to provide access control and privacy protection to Blockchains. Public Key Infrastructure is a mechanism to authenticate users in which one or more trusted organizations give digital signatures to users. Each signature is a pair of cryptographic keys (called public/private keys), which certifies that it belongs to a specific user. This means the keys can be used to encrypt health records: Li \textit{et al.}~\cite{Li_2019} describe an intermediate layer that validates all access requests and responds with encrypted messages encoded using users' public keys.

The RBAC mechanism uses ``roles'' and predefined permissions to control access to records. In Mahore \textit{et al.}~\cite{Mahore_2019}, the roles are defined as the users of the system (such as patients, hospitals, and researchers) each one having different permissions to access the health records.

In addition to access control, there is another feature of privacy: data ownership, i.e. the individual's ability to control his/her information. Most selected studies are in favor of this condition, and recommend adopting patient-centric approaches to health information storage and exchange solutions. 

This SLR seeks to analyze the use of self-sovereign identities to empower patient data ownership. Dealing with personal health information in terms of someone's identity allows patients to make claims about their health status that can be verified without disclosing confidential information. Houtan \textit{et al.}~\cite{Houtan_2020} analyze this trend and examine the potential of Blockchain technologies to enable PHR solutions by using self-sovereign identities. Buzachis \textit{et al.}~\cite{Buzachis_2019} employ uPort~\cite{Uport_2020}, a self-sovereign identity platform, to establish patients' identification. Zhou \textit{et al.}~\cite{Zhou_2018} provide a distributed data vending system for personal healthcare information, and employ similar concepts of sovereign identity claims to create a marketplace for healthcare data.

%RQ4
\subsection{RQ4:How does the current research regard the data ownership and access control aspects of sharing electronic health records? \newline RQ4.1: Who can add health records to the Blockchain? \newline RQ4.2: Who can read the health records obtained from the Blockchain?}

The RQ4 focuses on examining factors regarding the information exchange aspects of electronic health records by describing the scenarios where this exchange occurs. The RQ 4.1 and RQ4.2 seek to distinguish between the engaged parties and the roles they play in this exchange.

\subsubsection{Results}

The capabilities represented by \ref{table:capabilities} characterize the different healthcare stakeholders that participate in the EHR scenarios. The list below describes these entities:
\begin{itemize}
    \item \textbf{Patients:} the people whom health data concern. 
    \item \textbf{Physicians:} specialist doctors trained to diagnose patients
    \item \textbf{Hospitals:} places where patients go for medical treatment.
    \item \textbf{Diagnostic labs:} places that collect patients' biomedical information and provide health reports
    \item \textbf{Researchers:} academic professionals concerned with collecting anonymous medical information to conduct their research.
    \item \textbf{Insurance companies:} entities that offer medical insurance coverage for patients
\end{itemize}

The selected studies full-text review of the selected studies also made it possible to distinguish between three distinct roles for conducting electronic health data exchange:
\begin{itemize}
    \item Entities that \textbf{create healthcare data};
    \item Entities that \textbf{have access to healthcare data};
    \item Entities that \textbf{own the healthcare data}
\end{itemize}

Given the EHR scenarios mapped in Table \ref{table:capabilities}, healthcare entities can be classified in terms of the roles they play in each scenario. Figure \ref{fig:healthcarematrix} shows the selected papers and the relationship between healthcare entities and the roles they play.

\subsubsection{Analysis and discussion}

\begin{figure*}[ht]
    \centering
    \includegraphics[width=0.98\textwidth]{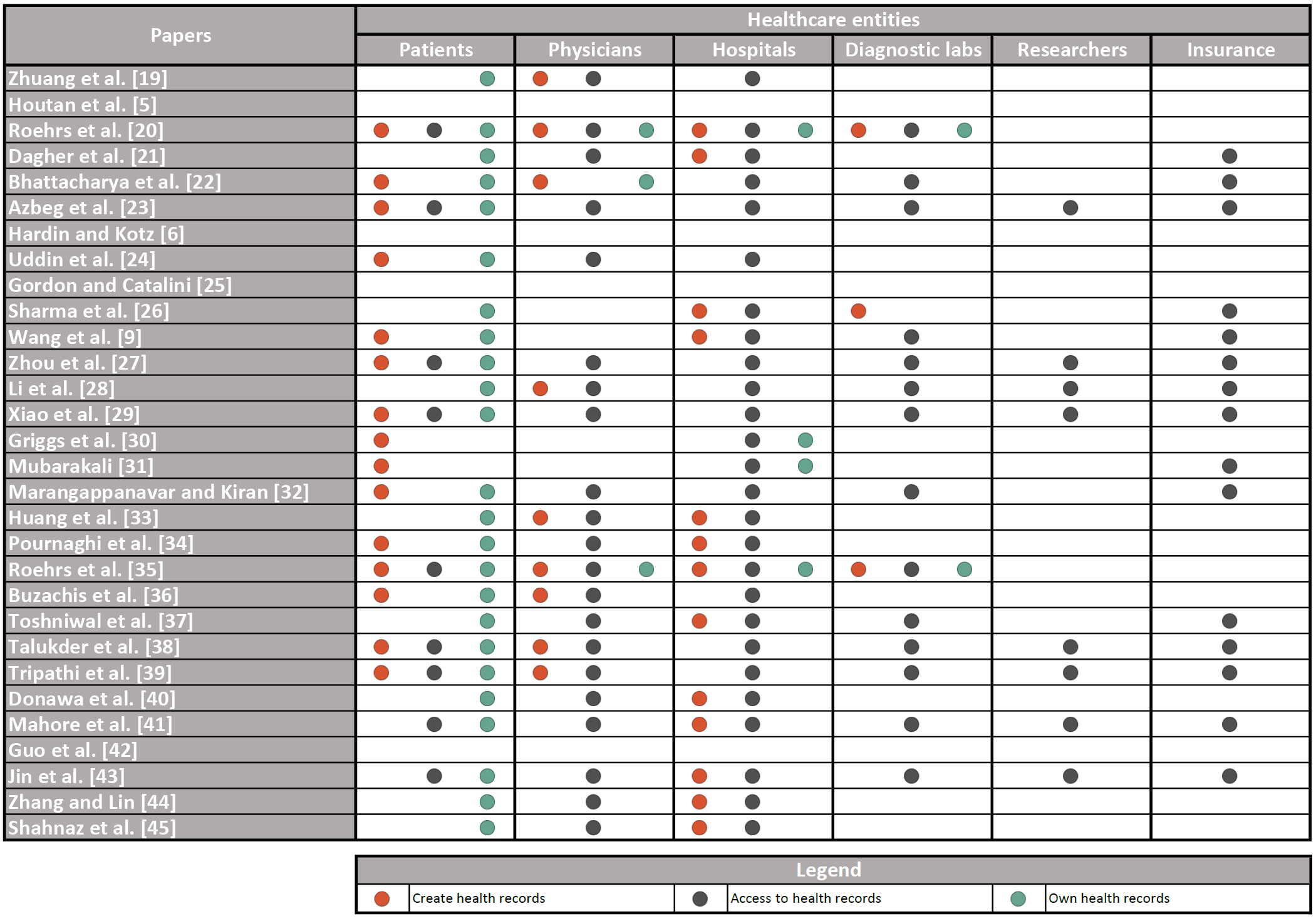}
    \caption{Classification of studies by health entities together with their performed roles}
    \label{fig:healthcarematrix}
\end{figure*}

Regarding self-sovereign identity principles, the matrix in Figure \ref{fig:healthcarematrix} shows misconceptions regarding data ownership. Most studies endorse the idea of giving patients the prerogative to control their own medical information. Nevertheless, this data ownership only emerges in data production and consent granting capabilities: 83.3\% of studies claim that patients are the owners of their health data, but only 30\% consider the question of whether patients should have access to their health records.

The Patient-centric approaches appear to focus on making health data available to other parties, and granting the patients access to their data, primarily hospitals (87.0\%) and doctors/physicians (70.0\%), followed by insurance companies (50.0\%), diagnostic labs (47.0\%) and researchers (47.0\%). The ability to collect health data straight from patients is also taken into account: 50\% of the studies describe patient-generated health data, either through wearable devices (33\%) like smartwatches and wellness bands or medical monitoring sensors (17\%).

Some studies explore ways of making patients' health information available for scientific research. These studies either suggest the use of public Blockchains containing anonymous data~\cite{Zhou_2018},~\cite{Talukder_2018},~\cite{Tripathi_2020} or private/permissioned Blockchains which enable patients to grant researchers access to their medical information~\cite{Bhattacharya_2020, Azbeg_2018, Li_2019, Xiao_2018, Mahore_2019, Jin_2019}.

\section{Challenges and Future Research}
\label{sec:challenges}

In this section, we discuss the challenges raised by the use of Blockchains and other open research questions such as the application of SSI to the healthcare domain.

\subsection{Challenges}

\textit{Interoperability and trusting in real-time and virtual interactions:} In the near future, several Blockchains and Self-Sovereign Identity systems will arise for both the same, and different, domain applications. These systems will be managed by private and public health institutions, banks and virtual financial platforms (e.g. for exchanging cryptocurrency), insurance companies, among several others. In this situation, each person will have many identities, issued by each system he/she interacts with, and it will be necessary to have some reliable way of obtaining these identities, while maintaining anonymity. When obtaining these identities, it is not enough just to trust in the person's storage device (e.g. wallet), because some of them may be hidden or lost, nor can someone rely on a trusted centralized organization, because the sovereignty of his/her information might be lost. It should be noted that, even if the person gives some identity in a real face-to-face interaction (e.g., when buying a house with a cryptocurrency), this situation can be a problem if this identity refers to an anonymous entity in a virtual interaction (e.g. in a cryptocurrency platform). Assuming that key identities can be obtained, there are still some open issues that must be addressed, because each Blockchain might have its own access permission rules, and data exchange format, among many other interoperability concerns. In light of this, it is necessary to adopt standards to account for these interactions and interoperability problems, that are being addressed in part by the W3C Verifiable Credentials Working Group and others similar organizations.

\textit{User adoption:} Today, smartphones are ubiquitous and an intrinsic part of people's everyday lives. Since these devices are constantly increasing their processing power, data transfer rate, and storage capacity, it might be a good idea to use them to form a part of the Blockchain architecture. Before this can happen, we must take note of some human and technical factors. In the first place, it is necessary to encourage users to use the smartphones not just as a wallet repository, as is currently the case, but as a peer. In that scenario, the device will consume its resources for mining the transactions, creating a consensus, storing the chain, and so forth. Most of the Blockchain systems and cryptocurrency platforms are based on financial rewards to give users incentives, but this kind of reward cannot be applied to other domains. In the second case, assuming that some sort of incentive is created, there will be several technological features that need to be addressed in order to make use of these devices. For example, it is necessary to consider that they have different capabilities and are subject to restrictions, such as Internet intermittency, reduced data transfer rates and limited storage space. Moreover, the battery consumption could lead to the system being turned off, among others problems.

\textit{Blockchain as a Service:} As mentioned in this review, SSI systems are combined with Blockchains and share its properties of immutability, replication, and anonymity, among other factors. Through this relationship, identities created by SSI systems can offer verifiable claims in a trusted way. In light of this, companies must provide Blockchain technologies with services that allow users to customize or use them to create new business applications. These services, which are specifically related to Blockchain, can vary from the low-level to high-level requirements, and cater for different kinds of users. For example, in the low-level requirements, software architects could choose either to use one ledger per user, or one ledger for all users, or even if the ledger has to be integrated with some Big Data analytic platform. In the high-level requirements, business users can easily create warning systems when some criteria is met, and perhaps analyzed with the Big Data platform chosen by the architect.

\textit{Usability and user experience:} Since Blockchain applied to the health domain will mainly be used by non-IT people (patients, doctors/physicians, and nurses, among others), it is necessary to devise a useful system that conceals all the complicated technological devices that support Blockchain. In view of this, all the mechanisms such as consensus, transfer data between on-chain and off-chain, certificate validation, etc., must be abstracted in such a way that, preferably, the user does not know about how these mechanisms work. User experience guidelines could be used to ensure that users can interact with the system without difficulty.

\textit{Sponsorship and Regulation:} As mentioned in this review, SSI can give a person the ability to manage his/her identities and share them with someone else. However, it should be noted that some of these identities must be issued by a trusted authority, to prevent malicious people or organizations from creating and using fake identities. For example, a person should be forbidden from self-generating digital claims. In a health domain context, we think that third parties (such as a government institutions, pharmaceutical companies, etc.) could act as a SSI sponsor that could influence the SSI dissemination and use among people through its brand name or reputation.

\textit{Security attacks and mitigation:} As mentioned when discussing the problem of usability, several non-IT people will use Blockchain-based systems. In the short term, they tend use their cellphones for managing and checking identity claims between each other. It is thus necessary to focus on all the security issues that could arise when someone abuses the system. A number of well-used techniques could be applied at the application layer the health domain, such as two-factor authentication, encrypted wallet backup and recovery, among others. On the other hand, there are non-application layer attacks that could affect the security of the user's device and, hence, the system itself. For example, some hardware or operating system bug can allow the capture of unencrypted messages (directly at the hardware level), that could be abused by malicious user. As well as mitigating cyber attacks to user's device, account should be taken of attacks against the Blockchain infrastructure (e.g, the peers storing the chain or abusing a smart contract bug). This could threaten the security of the components at different levels, such as by allowing the creation of a fake smart contract that phishes real user identities, or the creation of malicious ledgers to validate fake claims, among others risks. 

\subsection{Research opportunities}

\textit{Smart-contract Generator:} Some situations require that access to health information is restricted to specific health professionals. For example, some EHRs store both the anamnesis results and images of a patient: the former is accessible by any doctor/physician, while only cardiologists can access the latter. It is necessary to create a health domain service that observes the user's access consents and automatically generates and deploys a formally verified smart contract. This smart contract could manage several aspects of the EHR life cycle, such as the following: control over who can view, alter, store, or discard  EHRs, or who can retrieve the EHR from the storage location, and manage the insertion/deletion of new permissions, among others tasks.

\textit{Scalable platform for IoT support:} Currently, billions of IoT devices are used to monitor different medical scenarios. These applications monitor vital signs like heart rate, sleep patterns, stress factors, and daily step count. As shown in several studies, a group of IoT devices can generate millions of events per second. Thus, it is necessary to create a scalable Blockchain architecture that can register, store, and analyze this massive data amount. For limited resource devices (e.g., smartwatches), the scalability must account for both the throughput (i.e., the events processed per second) and the payload size.

% \textit{Use case validation}.

%\textit{Killer application:} \color[red]{We think that an easy way to increase the SSI adoption is to design and deploy a killer %application that is attractive to end-users. Several factors could influence this decision: it might cater for real needs or %rewards might be very attractive; because it is mandatory, etc. One example could be a COVID-19 application, that allows a %company worker to show that he/she has been immunized against the disease, which could improve public health decisions at %different levels. At the company level, this allow the worker to stay at work without any risk to his health or those of his %colleagues. At the city level, consolidated anonymous information will allow the stakeholders to know if a herd immunization was %achieved. At the State or country level, this will help determine how many vaccines must be bought. In addition, the proof of %immunization could be a requirement to obtain social welfare benefits}.

\textit{Self-sovereign health registry systems:} Self-sovereign identity systems offer an innovative way to handle digital identities, storing people's credentials in personal digital wallets. Those identities can be presented as verifiable credentials that can be proven using cryptographic keys stored in decentralized ledgers like Blockchain networks, dismissing the need for a central authority.
People have biometric patterns in the same way that they carry health markers. To take the identity/health features analogy further, diagnostic labs can issue blood test results for an individual in the same way a state agency issues a driver's license: as a set of claims about a person. Handling health information as a claim of someone's digital identity enables leveraging self-sovereign identity concepts to create decentralized electronic health record systems that give patients complete control over their health information. 
The authors of this systematic literature review intend to continue their research by applying the obtained results to design and build a self-sovereign health registry system prototype. This proof of concept aims to develop a personal health record system where medical service providers can store diagnostic test results, physician appointments, vaccination records, and other health information regarding the patient in his/her digital wallet. The patient can share his medical history directly with other health service providers without the need for a central repository or central authority, ensuring the patient's privacy and ownership over his/her data.

\section{Conclusion}
\label{sec:conclusions}

Electronic health records solutions are abundant in the market, and their numbers are still growing. Academic research is leaning towards these themes and exploring new technologies and strategies to address them, motivated by industrial regulations, automation processes, paperless initiatives, privacy and security risk mitigation, and the need to make an improvement in the health and wellness of patients. 
This study contributed to this debate by comprehensively analyzing key studies regarding SSI and Blockchain technologies. Below, a summary of the conclusions:

\hfill

\noindent
\textbf{It is still a novel subject: } medical institutions have been exploring this matter for some time, but the urge to break the silos of health information, (mainly driven by privacy concerns and the proliferation of Internet of Medical Things - IoMT devices) is requiring decentralized and interoperable solutions. The number of studies obtained from digital library databases increased from dozens in 2017 to hundreds in 2020, which shows the extent of academic interest in the subject.

\hfill

\noindent
\textbf{Blockchain technologies provide a viable means of addressing the fundamental challenges of EHR solutions:} electronic health records face a number of obstacles to their adoption, such as the following: access control, data integrity, interoperability, and auditing. Distributed Ledger Technologies like Blockchain solved similar problems in many other industries, which makes them a suitable alternative for healthcare data management.

\hfill

\noindent
\textbf{Most selected studies adopt consolidated Blockchain technologies as core components of their strategies:} Ethereum and Hyperledger are robust platforms that support many Blockchain systems in all kinds of industry and are at the heart of most studies reviewed by this SLR. Only studies that require definite outcomes recommend using custom implementations of Distributed Ledger Technologies.

\hfill

\noindent
\textbf{Keeping raw health data in external repositories (off-chain) can enable scalable solutions to be found:} just storing health record metadata and pointers to external raw files in the Blockchain can help deliver feature-rich and scalable designs.

\hfill

\noindent
\textbf{The adoption of health data standards to store information on-chain allows users to easily filter/select health records from Blockchain data structures more easily:} the health industry has standardized data like OpenEHR, ISO 13606 and HL7 FHIR to enable interoperability between EHR systems. Compliance with these standards when storing patient information in the Blockchain provides client application with the means to search for particular records without accessing the raw information. This feature simplifies data management and helps ensure patients privacy.

\hfill

\noindent
\textbf{Smart contracts are an essential part of Blockchain-based electronic health records systems:} reviewed studies suggest that an intermediary (proxy) layer, acting as a controller between the client applications and the Blockchain, can perform critical functions with regard to access control, cryptography, consent management, and data classification. Smart contracts are the most versatile technology for achieving these outcomes.

\hfill

\noindent
\textbf{The data ownership concept should be expanded to ensure that patients have complete control over their data:} a number of studies have found patient-centric EHR solutions regarding the ability of patients to grant or revoke access to their health records and this are enough for this to be considered data ownership. This concept is misleading because patients should be able to have full control over their data. Adopting predicates from the principles of self-sovereign identities (SSI) could make patient-centric solutions more accurate.

\bibliographystyle{ACM-Reference-Format}
\bibliography{references}

\newpage

\appendix
\section{List of papers and quality scores}

\begin{table}[h]
\tiny
\renewcommand{\arraystretch}{1.2}
\caption{List of the selected papers for this SLR along with their quality scores and citation count}
\label{table:listofselectpapersqa}
\centering
\begin{tabular}{ p{0.5cm} | r | r | r | r | r | r | r | r | r | r | r | r | r | r}
 \hline
 Ref. & QA1 & QA2 & QA3 & QA4 & QA5 & QA6 & QA7 & QA8 & QA9 & QA10 & QA11 & QA12 & Score & Citations \\ 
 \hline
\cite{Zhuang_2020} & 1 & 1 & 0 & 0 & 0 & 0 & 1 & 0 & 1 & 1 & 1 & 1 &  7.0  & 1\\
\cite{Houtan_2020} & 1 & 1 & 1 & 0 & 1 & 0 & 0 & 0.5 & 1 & 1 & 1 & 1 &  8.5  & 0\\
\cite{Roehrs_2019} & 0.5 & 0.5 & 1 & 0 & 0 & 0 & 1 & 1 & 1 & 1 & 1 & 1 &  8.0  & 24\\
\cite{Dagher_2018} & 1 & 1 & 0 & 0 & 0 & 0 & 1 & 1 & 1 & 1 & 1 & 1 &  8.0  & 159\\
\cite{Bhattacharya_2020} & 1 & 1 & 1 & 0 & 0 & 0 & 1 & 1 & 1 & 1 & 1 & 1 &  9.0  & 19\\
\cite{Azbeg_2018} & 1 & 1 & 1 & 0 & 0 & 0 & 1 & 0 & 1 & 0.5 & 1 & 0.5 &  7.0  & 4\\
\cite{Hardin_2019} & 1 & 1 & 1 & 0 & 0 & 0 & 0.5 & 0.5 & 1 & 1 & 1 & 1 &  8.0  & 2\\
\cite{Uddin_2020} & 1 & 1 & 1 & 0 & 0 & 0 & 1 & 1 & 1 & 1 & 1 & 1 &  9.0  & 5\\
\cite{Gordon_2018} & 1 & 1 & 1 & 0 & 0 & 0 & 0 & 0.5 & 1 & 1 & 1 & 1 &  7.5  & 191\\
\cite{Sharma_2020} & 1 & 1 & 0 & 0 & 0 & 0 & 1 & 0 & 1 & 1 & 1 & 1 &  7.0  & 0\\
\cite{Wang_2019} & 1 & 1 & 0 & 0.5 & 0 & 0 & 1 & 1 & 1 & 1 & 1 & 1 &  8.5  & 7\\
\cite{Zhou_2018} & 1 & 1 & 0 & 0 & 0 & 1 & 0 & 1 & 1 & 1 & 1 & 1 &  8.0  & 6\\
\cite{Li_2019} & 1 & 1 & 0.5 & 0.5 & 0 & 0 & 1 & 0 & 1 & 0.5 & 1 & 1 &  7.5  & 16\\
\cite{Xiao_2018} & 1 & 1 & 0 & 0.5 & 0 & 0 & 1 & 0 & 1 & 1 & 1 & 1 &  7.5  & 7\\
\cite{Griggs_2018} & 0.5 & 1 & 1 & 0 & 0 & 0 & 1 & 1 & 1 & 1 & 1 & 1 &  8.5  & 167\\
\cite{Mubarakali_2020} & 1 & 1 & 1 & 0 & 0 & 0 & 1 & 1 & 1 & 1 & 1 & 1 &  9.0  & 1\\
\cite{Marangappanavar_2020} & 1 & 1 & 0 & 0 & 0 & 0 & 1 & 1 & 1 & 1 & 1 & 1 &  8.0  & 0\\
\cite{Huang_2019} & 1 & 1 & 0 & 0 & 0 & 0 & 1 & 0 & 1 & 1 & 1 & 1 &  7.0  & 0\\
\cite{Pournaghi_2020} & 1 & 1 & 0 & 0 & 0 & 0 & 1 & 1 & 1 & 1 & 1 & 1 &  8.0  & 4\\
\cite{Roehrs_2017} & 1 & 1 & 1 & 0 & 0 & 0 & 1 & 0 & 1 & 1 & 1 & 1 &  8.0  & 119\\
\cite{Buzachis_2019} & 1 & 1 & 1 & 0 & 1 & 0 & 1 & 0 & 1 & 0.5 & 1 & 1 &  8.5  & 0\\
\cite{Toshniwal_2019} & 1 & 1 & 0 & 0 & 0 & 0 & 1 & 0 & 1 & 1 & 1 & 1 &  7.0  & 1\\
\cite{Talukder_2018} & 1 & 0.5 & 1 & 1 & 0 & 0 & 1 & 0 & 1 & 1 & 1 & 0.5 &  8.0  & 12\\
\cite{Tripathi_2020} & 1 & 1 & 1 & 0 & 0 & 0 & 1 & 0 & 1 & 1 & 1 & 1 &  8.0  & 5\\
\cite{Donawa_2019} & 1 & 1 & 0 & 1 & 0 & 0 & 1 & 1 & 1 & 1 & 1 & 1 &  9.0  & 0\\
\cite{Mahore_2019} & 1 & 1 & 0 & 1 & 0 & 0 & 1 & 1 & 1 & 1 & 1 & 1 &  9.0  & 0\\
\cite{Guo_2018} & 1 & 1 & 0 & 1 & 0 & 0 & 1 & 1 & 1 & 1 & 1 & 1 &  9.0  & 161\\
\cite{Jin_2019} & 1 & 1 & 0 & 1 & 0 & 0 & 1 & 0 & 1 & 1 & 1 & 1 &  8.0  & 0\\
\cite{Zhang_2018} & 1 & 1 & 0 & 0 & 0 & 0 & 1 & 1 & 1 & 1 & 1 & 1 &  8.0  & 82\\
\cite{Shahnaz_2019} & 1 & 1 & 0 & 0 & 0 & 0 & 0 & 1 & 1 & 1 & 1 & 1 &  7.0  & 19\\
\hline
\end{tabular}
\end{table}

\end{document}